\documentclass[preprint,review,12pt]{elsarticle}
\usepackage{graphicx}
\usepackage{epsf}
\usepackage{amsmath}
\usepackage{lscape}
\usepackage{color}
\usepackage{dcolumn}
\usepackage{hyperref}
\usepackage{amsmath}
\usepackage{rotating}
\usepackage{soul}
\usepackage{longtable}
\usepackage{natbib}

\newcommand{\cm}{cm$^{-1}$}
\newcommand{\m}{$\mu$m}
\newcommand{\p}{$^\prime$}
\newcommand{\pp}{$^{\prime\prime}$}

\newcommand{\NH}{NH$_{3}$}

\newcommand{\abinitio}{\textit{ab initio }}

\journal{Journal of Molecular Spectroscopy}

\begin{document}
	
	\begin{frontmatter}
		
		\title{A near infrared line list for \NH: Analysis of a Kitt Peak spectrum after 35 years}
		\author{Emma~J. Barton, Sergei.~N. Yurchenko, Jonathan Tennyson}
		\address{Department of Physics and Astronomy, University College London,
			London, WC1E 6BT, UK}
		\author{S. B\'{e}guier, A. Campargue}
		\address{Universit\'{e} Grenoble Alpes, LIPhy, F-38000 Grenoble, France.}
		\address{CNRS, LIPhy, F-38000 Grenoble, France} 
		
		\begin{abstract}
			
			A Fourier Transform (FT) absorption spectrum of room temperature 
			\NH\ in the region 7400 - 8600 \cm\ is analysed using a 
			variational line list and ground state energies determined using 
			the MARVEL procedure. The spectrum was measured by Dr Catherine 
			de Bergh in 1980 and is available from the Kitt Peak data center. 
			The centers and intensities of 8468 ammonia lines were retrieved 
			using a multiline fitting procedure. 2474 
			lines are assigned to 21 bands providing 1692
			experimental energies in the range 7000 - 9000 \cm. The spectrum 
			was assigned by the joint use of the BYTe variational line list 
			and combination differences. The assignments and experimental 
			energies presented in this work are the first for ammonia in the 
			region 7400 - 8600 \cm, considerably extending the range of known 
			vibrational-excited states.

		\end{abstract}
		
	\end{frontmatter}
	
	\section{Introduction}
	
	Being one of the simplest polyatomic molecules and present in many 
	environments, including the interstellar medium, brown dwarfs and 
	solar system planets, \NH\ is a very important molecule for 
	astronomers. It also has several applications in industry, such as 
	the reduction of NOx emissions in smoke stacks \cite{11DTxxxx.nh3} 
	and the manufacture of hydrogen cyanide by the Andrussow process 
	\cite{58Pixxxx.nh3}. This has motivated over 140 experimental 
	studies on its spectrum, recent work includes high 
	\cite{jt508,12HaLiBea,12HaLiBeb,jt616} and low 
	\cite{14CeHoVeCa,14CaCeCoRo} temperature studies in various 
	spectral regions. A comprehensive compilation of measured \NH\ 
	rotational and ro-vibrational spectra can be found in a recent 
	MARVEL study \cite{jt608}. 
	
	The MARVEL (measured active rotation-vibration energy levels) 
	algorithm \cite{jt412,12FuCsi.method} simultaneously analyses all 
	available assigned and labelled experimental lines, thus yielding 
	the associated energy levels. The recent study for \NH\ analysed 
	29,450 measured transitions and yielded 4961 accurately-determined 
	energy levels. The critically reviewed and validated high 
	resolution experiments employed by this study, cover the region 
	0.7 - 17,000 \cm\ with a large gap between 7000 -- 15,000 \cm. In 
	fact there is an overall lack of detailed and accurate information 
	for \NH\ transitions in this region.
	
	The band model parameters of Irwin {\it et al} \cite{99IrCaSi.nh3} 
	cover the region 400 to 11000 \cm, but was intended for analysis 
	at low spectral resolution so the measurements were obtained at a 
	spectral resolution of only 0.25 \cm\ and not assigned. The HITRAN 
	database \cite{jt557}, a major source of experimental data, contains 
	no information for \NH\ above 7000 \cm. 
	
	A number of variational line lists are available for \NH\  
	\cite{jt466,jt500,11HuScLe2.NH3}. 
	In this work we use BYTe \cite{jt500} which is a variationally 
	computed line list for hot \NH\ that covers the range 0 - 12,000 \cm. 
	BYTe is expected to be fairly accurate for all temperatures up to 
	1500 K (1226 $^{\circ}$C). It comprises of 1~138~323~251 transitions 
	constructed from 1~373~897 energy levels lying below 18~000 \cm. It 
	was computed using the NH3-2010 potential energy surface \cite{jt503}, 
	the TROVE ro-vibrational computer program \cite{07YuThJe.method} and 
	an \abinitio dipole moment surface \cite{jt466}. However this line 
	list is known to be less accurate for higher wavenumber transitions 
	\cite{11HuScLe,11HuScLe2,12SuBrHuSc}, and assigned high resolution 
	laboratory spectra in poorly characterised regions is needed. 
	
	The reason for the void between 7000 and 15,000 \cm\ is the 
	complexity of the \NH\ spectrum making analysis of experimental 
	spectra using the established method of fitting Hamiltonians tricky 
	\cite{95KlTaBr.NH3,99KlBrTa.NH3}. In the present work we take on this 
	challenge by employing the same technique used previously 
	\cite{jt508,jt616} to study high temperature spectra, to study the 
	room temperature, near infrared spectrum of \NH\ in the 7400 -- 8600 \cm\ 
	region. This region is of present interest. For example there are 
	peaks in \NH\ opacity between 1.210 \m\ and 1.276 \m\ which are 
	important features in late type T dwarfs \cite{jt596}. In wavenumbers 
	this region is 7836 - 8265 \cm\ which  is covered by an unanalysed 
	1980 room temperature spectrum in the Kitt Peak Archive. This spectrum 
	was recently used by Campargue {\it et al} \cite{15CaMiLoKa} to identify 
	residual \NH\ lines in their ultra-long pathlength water spectra.
	
	This article has the following structure. Section 2 describes the Kitt 
	Peak spectrum and the construction of the experimental line list. 
	Section 3 gives an overview of the assignment procedure. Section 4 
	comes in two parts. The accuracy of BYTe is assessed 
	in Section 4.1 by a direct comparison with the experimental line list. 
	A summary of all assignments and new experimental 
	energies is presented in Section 4.2. Finally Section 5 gives 
	our conclusions and discusses avenues for further work.   
	
	\section{Experimental Data}
	
	\subsection{The Kitt Peak spectrum}
	
\begin{figure}[htbp]
		\centering
		\scalebox{0.3}{\includegraphics{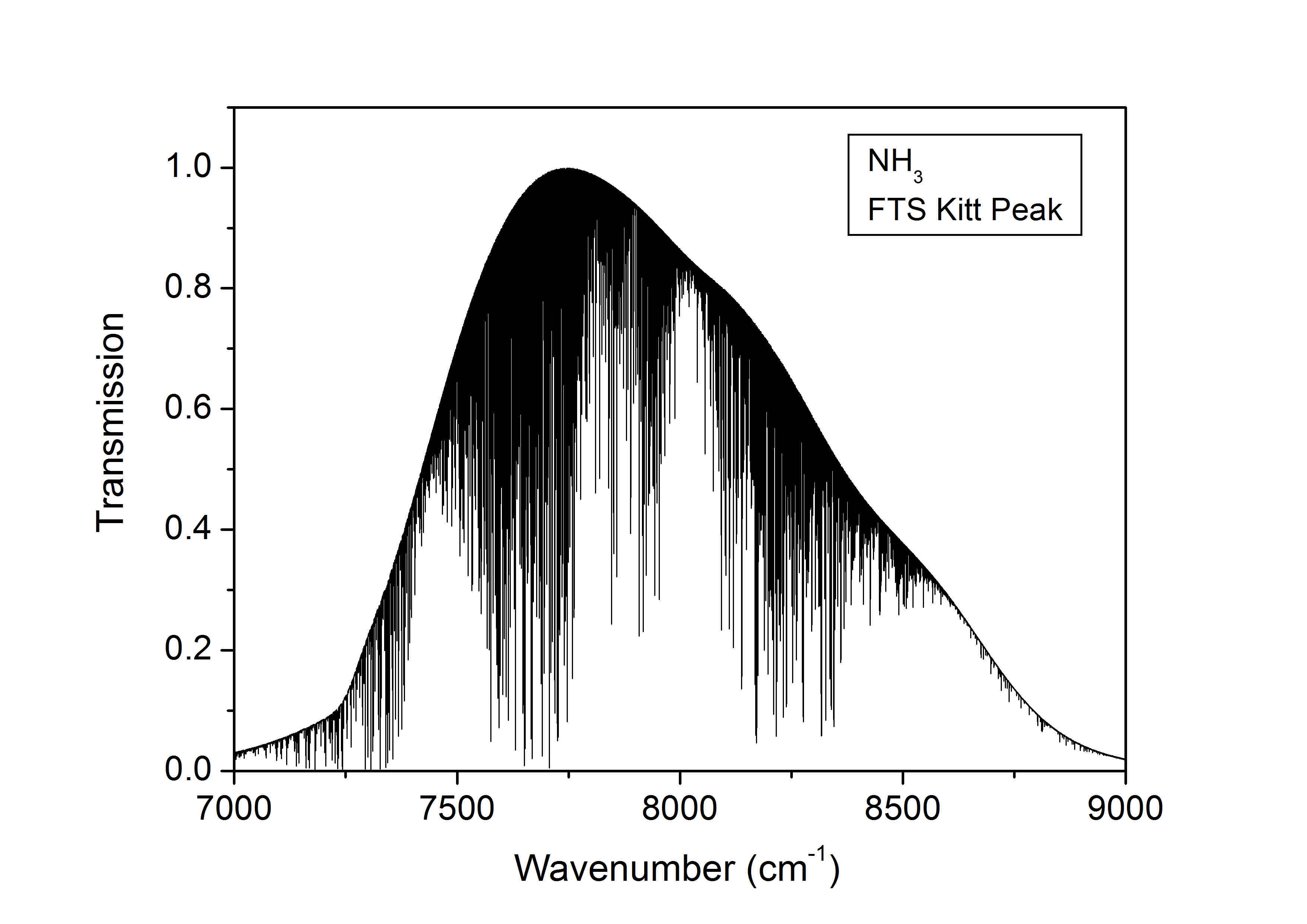}}
		\caption{Overview of Kitt Peak FT spectrum 800407R0.004.}
		\label{fig:overview}
	\end{figure}

	The Kitt Peak data center provides open access to laboratory 
	Fourier Transform (FT) spectra recorded at Kitt Peak. 
	The room temperature laboratory absorption spectrum of \NH\ 
	analysed by the present work (800407R0.004) was recorded by 
	Dr. Catherine de Bergh using a one metre FT spectrometer. 
	The spectrometer in question was a permanent instrument on the 
	McMath Solar Telescope, the largest solar telescope in the world 
	\cite{KPMcMath}, and was used for both solar and laboratory analysis. 
	In 2012 the instrument was transferred to Old Dominion University 
	\cite{KPMcMath2}. 
	The spectrum was recorded at a resolution of 0.01 \cm\ and generated 
	from an average of 12 scans. Some key information provided in the FITS 
	header is presented in Table~\ref{tab:fits}. The first and last wavenumber 
	are listed as ~ 5797 \cm\ and 9682 \cm\ respectively but our study focusses 
	on the region 7400 - 8600 \cm.  Figure~\ref{fig:overview} gives an overview
of the spectrum.
	
	\begin{table}
		\caption{Key experimental information provided in the FITS header downloaded from 
			the Kitt Peak Archive.} 
		\begin{center}
			\begin{tabular}{ll}
				\hline
				Archive Name   &   800407R0.004 \\
				Temperature    &  21.5 $^{\circ}$C   \\
				Pressure       &    5 Torr  \\
				Resolution     &   0.01109469 \cm\   \\
				Path length    &  25 m       \\
				Date           &    07/04/1980 \\
				Spectral Type  &    Absorption  \\
				Wavenumber Start &  5786.89  \\
				Wavenumber Stop  &  9682.65   \\
				\hline
			\end{tabular} \label{tab:fits}
		\end{center}
	\end{table}

	\subsection{Construction of the line list}
	
	The wavenumber scale was calibrated using water lines which 
	are clearly apparent around 7500 and 8600 \cm. The correction 
	term fitted as a linear function of the wavenumbers was found 
	to be -0.028 \cm\ at 8000 \cm. It is worth noting that the 
	shape of the water lines indicates that they involve a sharp 
	contribution due to water molecules present as an impurity in 
	the low pressure sample, and a broader contribution due to 
	atmospheric water present in the spectrometer. The quality of 
	the wavenumber calibration was checked in two ways: by 
	comparison of ammonia line positions observed at the low energy 
	range of the spectrum with positions provided in the HITRAN 
	database below 7000 \cm\, and by comparison to the positions of 
	ammonia lines identified in the high sensitivity CRDS spectrum 
	of water \cite{15CaMiLoKa}. As a result, the accuracy of the 
	reported line positions is estimated to be of the order of 
	$3\times 10^{-3}$ \cm\ for isolated lines of intermediate intensity. 
	
	The line centers and intensities were retrieved by multiline
        fitting of the spectrum using a homemade three step suite of
        programs written in Labview and C++ (see \cite{15BeLiCa.CH4}
        for details). First the transmittance spectrum was converted
        to absorbance neglecting the apparatus function

        After correction of the baseline, a peak finder procedure is
        used to construct a peak list (line center, peak height). A
        synthetic spectrum is then simulated by attaching a default
        profile to each peak. Taking into account the different
        factors limiting the accuracy of the intensity retrieval (line
        overlapping, baseline uncertainty, spectrometer apparatus
        function, significant line broadening), an effective default
        profile was determined from a small number of relatively
        isolated lines. Their profiles were fitted using a Voigt
        function whose averaged Gaussian and Lorentzian components
        were adopted as default values. From a visual comparison of
        the spectrum with the simulation, lines not detected by the
        peak finder were added manually. Then an automatic multiline
        fit was performed over the entire analysed region by adjusting
        only the line center and integrated line absorbance, the shape
        of all lines being fixed to the default Voigt profile.
        Finally, a manual adjustment was performed by further refining
        the profile parameters and adding/deleting weak lines. In the
        case of highly blended features, we tried to limit the number
        of components and preferred to relax the constraints on the
        parameter profiles within reason than to add lines in order to
        minimise the residuals.

\begin{figure}[htbp]
		\centering
		\includegraphics{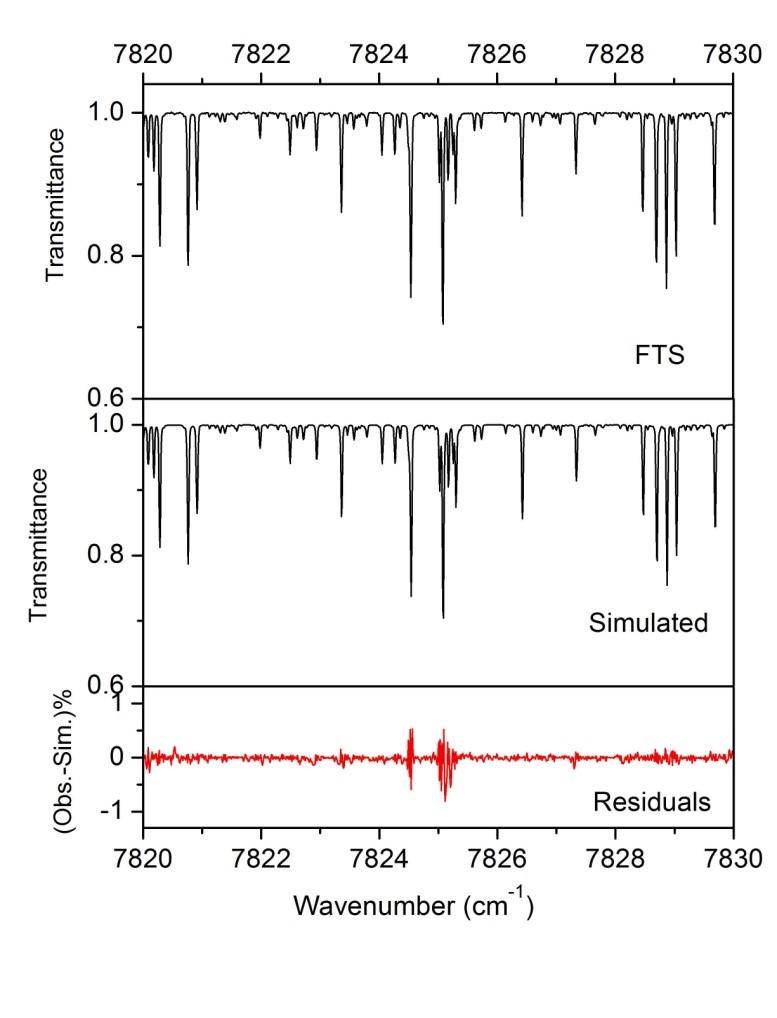}
		\caption{Illustration of the multiline fitting procedure adopted to retrieve the ammonia 
			line parameters. The FT spectrum (800407R0.004) was recorded by C. de Bergh 
			and corresponds to a 25 m pathlength and a 5 Torr pressure.}
		\label{fig:fitted}
	\end{figure}

        The apparatus function (neglected in the above tratement)
        introduces a significant underestimate of the intensity of
        the lines with transmittance smaller than
        30 \%. On the basis of spectral simulations, a correction curve
        was constructed for the line intensities; this
curve was validated using
        lines near 7000 \cm\ with intensity provided in the HITRAN
           database and applied to all the intensities. The
          correction is on the order of 20 \%\ for line intensities of
           $1\times10^{-22}$ cm/molecule.
 Figure~\ref{fig:fitted} illustrates
        the quality of the final spectral reproduction achieved.
	
\begin{figure}[htbp]
		\centering
		\scalebox{0.4}{\includegraphics{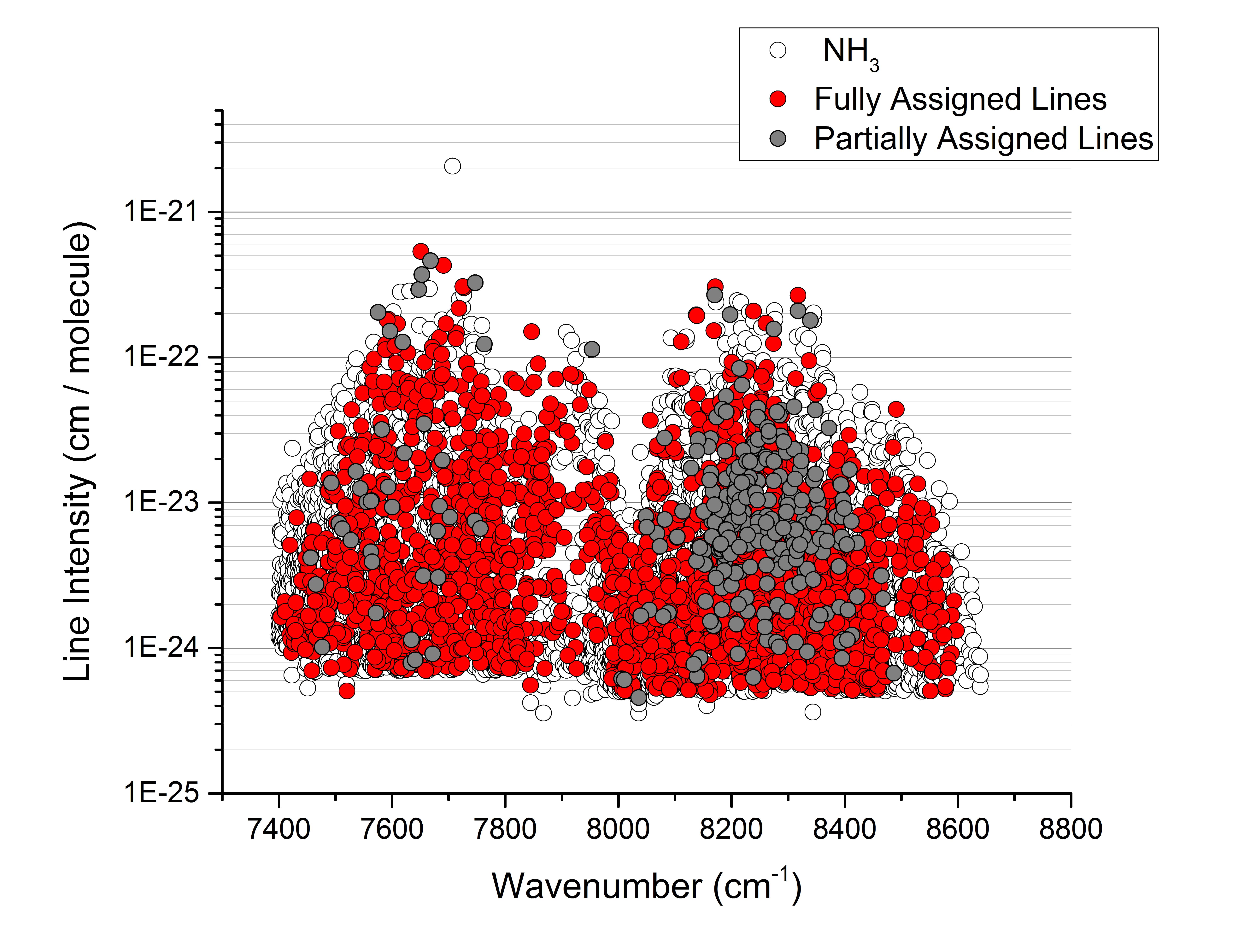}}
		\caption{Overview of the \NH\ line list retrieved between 7400 and 8640 \cm. 
			Fully assigned lines are highlighted 
			in red, partially assigned lines are shaded in grey. Please refer to 
			Section 4.2 for details.}
		\label{fig:explist}
	\end{figure}

	After removal of some water lines located at the low and high
        energy borders of the 7400 - 8640 \cm\ studied spectral
        region, the overall list presented in Fig.~\ref{fig:explist}
        includes 8468 lines with intensities ranging between about
        $5\times 10^{-25}$ and $5\times 10^{-22}$ cm/molecule. We
        suspect the retrieved intensity for the strongest experimental
        line (I $>$ $1\times 10^{-21}$ cm/molecule) is at least an
        order of magnitude too strong. This is because the strongest
        BYTe lines are all of the order $1\times 10^{-22}$ cm/molecule
        which is consistent with the remaining strongest experimental
        lines

        Fully assigned lines and partially assigned lines are
        separately highlighted in Fig.~\ref{fig:explist}; details of
        this distinction are discussed in  Section 4.2. The average
        uncertainty on the retrieved line intensities is estimated to
        be of the order 15 \%\ for isolated lines of intermediate
        intensity. This uncertainty value is consistent with the
        agreement we obtained by comparison to HITRAN intensity values
        for a sample of lines located in the 6950 - 7000 \cm\ range.
        It is nevertheless important to clarify that for the small
        fraction of lines with intensities larger than $1\times
        10^{-22}$ cm/molecule, the uncertainty values may
        significantly exceed 15\%. For these lines, the transmission
        at the line center is of the order of only a few percent and
        biases related to the zero transmission or a not fully
        characterized apparatus function may be significant.
	
	\section{Data Analysis}
	
	This study employs the BYTe \cite{jt500} variational line list 
	and ground state energies determined using the MARVEL procedure \cite{jt608}.
	
	A list of observable BYTe lines for the experimental conditions was 
	compiled. This list included all lines with an absorption intensity 
	greater that $5\times 10^{-25}$ cm/molecule at 292.5 K (21.5 $^{\circ}$C).   
	
	\subsection{Assignment Procedure}
	
	Tentative assignments were given to the strongest experimental lines 
	by comparison with the strongest BYTe lines (see Figure~\ref{fig:byte2}). 
	Combination difference partners for these tentative assignments were 
	found in the list of observable BYTe lines and assigned to experimental 
	lines where possible. In this case the BYTe intensity should be within 
	55\%\ of the experimental intensity and have a residual 
	(observed minus calculated difference) up to about 3 \cm. Upper state 
	energies for resulting combination difference pairs (PQ, RQ or PR) or 
	triplets (PQR) were calculated by adding the experimental line positions 
	to the energy for the lower state associated with the transition. As all 
	bands considered in this work go to the ground state and are 
	limited to a rotational excitation of $J \leq 16$, these energies are known 
	to very high accuracy (of the order 10$^{-4}$ \cm) and are available from the recent 
	MARVEL study \cite{jt608}. If the calculated upper energies agreed within 0.006 \cm\ for 
	intermediate intensity unblended lines or 0.02 \cm\ for weak or blended lines 
	these assignments were confirmed. Table~\ref{tab:CD} gives an subset of 
	assignments confirmed by combination differences. In some cases the derived upper state 
	energies were supported by more than three transitions.  
	
	Confirmed combination difference pairs and triplets for the same vibrational 
	band provide an expected observed minus calculated (Obs. - Calc.) difference 
	for all lines in that band. This Obs. - Calc. difference was used to shift 
	the BYTe line positions for vibrational bands with at least 5 combination 
	difference pairs or triplets and hence make further assignments to 
	those bands. This is the method of branches \cite{jt205},
	which exploits the systematic behavior of variational calculations that
	give an overall band shift but largely preserve the rotational
	structure. Table~\ref{tab:Br} 
	gives a subset of assignments made using the method of branches.
	
	The final assignment list is a compilation of all combination
        difference (CD) and branch (Br) assignments. Due to the
        inaccuracy of BYTe in the region 7400 - 8600 \cm\ (see Section
        4.1) unconfirmed assignments made by simple line list
        comparison were discarded. Finally, we note that it is possible
to identify further transitions that can satisfy CD criteria; however
these transitions do not fall into clear bands so we have left these
lines for future analysis.
	
	\begin{table}
		\caption{A sample of the assigned transitions confirmed by 
			combination differences. Abbreviated $(v_1 v_2 v_3^{L_3} v_4^{L_4})^{i}$ 
			vibrational labels, described in Section 4.2, followed by rotational quantum 
			numbers $J$ and $K$ are given below, full quantum assignments are provided in 
			the supplementary data.} 
		\begin{center}
			\begin{tabular}{llrrll}
				\hline
				Obs.    &  Calc. \cite{jt500}  & Upper                                 & Lower       & Type  & Upper State Energy  \\
				\cm         &  \cm         & Quanta                                & Quanta      & P/Q/R & \cm                 \\ 
				\hline
7555.8782	&	7554.8133	&	$(v_{1}+v_{2}+v_{3}^{1})^{-}$ 4 2	&	$0^{+}$ 5 0	&	 P	&	7853.5211	\\
7588.2929	&	7587.2278	&	$(v_{1}+v_{2}+v_{3}^{1})^{-}$ 4 2	&	$0^{-}$ 4 3	&	 Q	&	7853.5204	\\
7687.4314	&	7686.3683	&	$(v_{1}+v_{2}+v_{3}^{1})^{-}$ 4 2	&	$0^{+}$ 3 0	&	 R	&	7853.5209	\\
7734.2821	&	7733.2204	&	$(v_{1}+v_{2}+v_{3}^{1})^{-}$ 4 2	&	$0^{-}$ 3 3	&	 R	&	7853.5205	\\
7766.8616	&	7765.7996	&	$(v_{1}+v_{2}+v_{3}^{1})^{-}$ 4 2	&	$0^{-}$ 5 3	&	 P	&	7853.5206	\\
&		&		&		&		&		\\
7710.9095	&	7708.0570	&	$(v_{2}+2v_{3}^{2})^{+}$ 6 1	&	$0^{+}$ 7 2	&	 P	&	8250.7546	\\
7837.5170	&	7834.6673	&	$(v_{2}+2v_{3}^{2})^{+}$ 6 1	&	$0^{-}$ 6 1	&	 Q	&	8250.7549	\\
7849.1070	&	7846.2574	&	$(v_{2}+2v_{3}^{2})^{+}$ 6 1	&	$0^{+}$ 6 2	&	 Q	&	8250.7551	\\
7967.8183	&	7964.9705	&	$(v_{2}+2v_{3}^{2})^{+}$ 6 1	&	$0^{+}$ 5 2	&	 R	&	8250.7555	\\
&		&		&		&		&		\\
7519.4722	&	7517.1594	&	$(v_{1}+v_{2}+2v_{4}^{2})^{+}$ 6 5	&	$0^{-}$ 7 7	&	 P	&	7894.5792	\\
7537.0004	&	7534.6874	&	$(v_{1}+v_{2}+2v_{4}^{2})^{+}$ 6 5	&	$0^{+}$ 6 4	&	 Q	&	7894.5846	\\
7655.9330	&	7653.6210	&	$(v_{1}+v_{2}+2v_{4}^{2})^{+}$ 6 5	&	$0^{+}$ 5 4	&	 R	&	7894.5856	\\
&		&		&		&		&		\\
8161.8067	&	8159.5253	&	$(v_{1}+3v_{4}^{1})^{-}$ 8 5	&	$0^{+}$ 9 7	&	 P	&	8872.6632	\\
8321.9021	&	8319.6317	&	$(v_{1}+3v_{4}^{1})^{-}$ 8 5	&	$0^{+}$ 7 1	&	 R	&	8872.6608	\\

				\hline
			\end{tabular} \label{tab:CD}
		\end{center}
	\end{table}
	
	\begin{table}
		\caption{A sample of assignments made to vibrational band $(v_{2} + 2v_{3}^{2})^{-}$ using 
			the method of branches. The expected Obs. - Calc. of ~ 2.4 \cm\ was determined by averaging 
			the residuals from 21 combination difference 
			pairs and 20 combination difference triplets.} 
		\begin{center}
			\begin{tabular}{llrrrrll}
				\hline
				Obs.    &  Calc. \cite{jt500} & J\p & K\p & J\pp & K\pp & Type  & Obs. - Calc. \\
				\cm         &  \cm        &     &     &      &      & P/Q/R & \cm          \\          
				\hline
7833.3464	&	7833.3464	&	1	&	1	&	2	&	2	&	P	&	2.3824	\\
7847.1197	&	7847.1197	&	0	&	0	&	1	&	1	&	P	&	2.3859	\\
7704.1322	&	7704.1322	&	10	&	10	&	11	&	11	&	P	&	2.3772	\\
7813.0536	&	7813.0536	&	8	&	5	&	8	&	4	&	Q	&	2.4110	\\
7822.9094	&	7822.9094	&	7	&	4	&	7	&	3	&	Q	&	2.4245	\\
7895.3264	&	7895.3264	&	10	&	8	&	10	&	9	&	Q	&	2.4227	\\
8008.9997	&	8008.9997	&	12	&	11	&	11	&	10	&	R	&	2.4134	\\
8010.4040	&	8010.4040	&	12	&	7	&	11	&	6	&	R	&	2.4159	\\
8000.4518	&	8000.4518	&	11	&	9	&	10	&	8	&	R	&	2.3839	\\   			
				\hline
			\end{tabular} \label{tab:Br}
		\end{center}
	\end{table}    
	
	\section{Results and Discussion}
	
	The accuracy of BYTe in the region 7400 - 8600 \cm\ is assessed by 
	comparison to the experimental line list. BYTe is then 
	employed to initiate assignments by combination differences and 
	the method of branches. The experimental line list (with full or partial 
	quantum labels for assigned lines) 
	and new energy level information derived from the assignments 
	are presented in supplementary data. 
	
	\subsection{Direct Comparison with BYTe}
	
\begin{figure}[htbp]
		\centering
		\scalebox{0.4}{\includegraphics{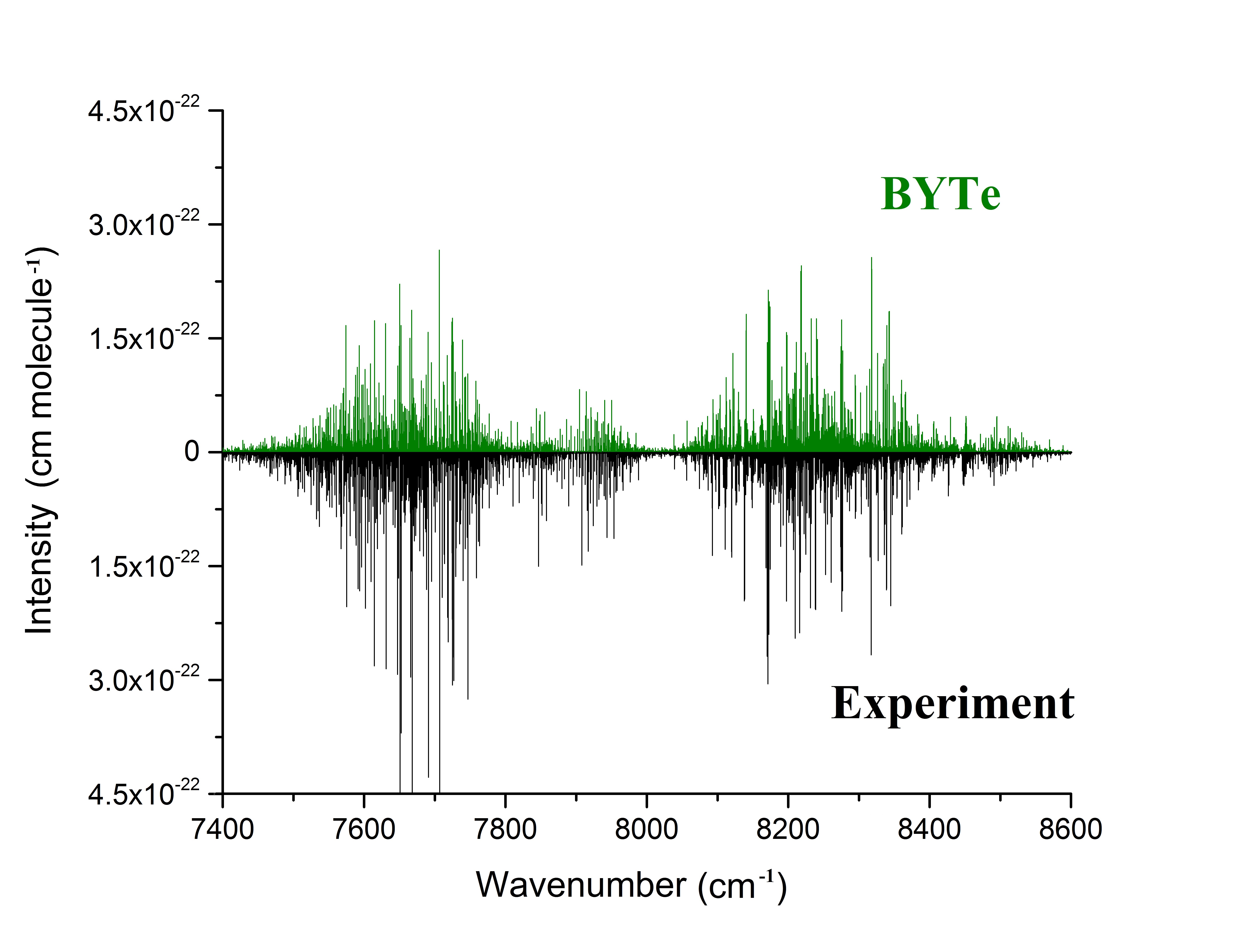}}
		\caption{Comparison between experimental (black) and calculated BYTe (green) stick spectra 
			at 21.5 $^{\circ}$C for the range 7400 - 8600 \cm.}
		\label{fig:byte1}
	\end{figure}

\begin{figure}[htbp]
		\centering
		\scalebox{0.4}{\includegraphics{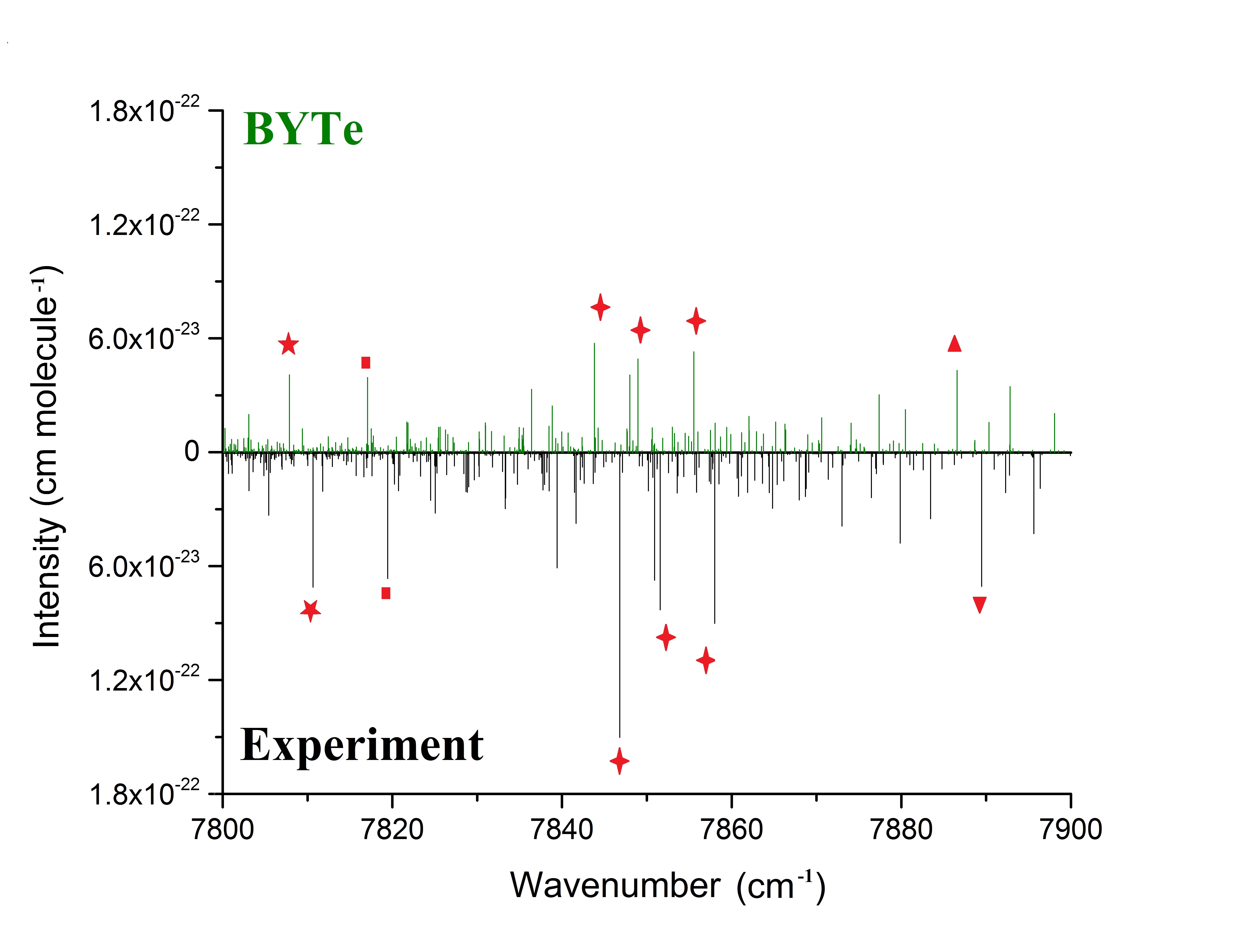}}
		\caption{Comparison between experimental (black) and calculated BYTe (green) stick spectra 
			at 21.5 $^{\circ}$C for the range 7800 - 7900 \cm. Symbols indicate examples 
			of tentative assignments made by line list comparison.}
		\label{fig:byte2}
	\end{figure}

	The BYTe variational line list covers the range 0 -- 12,000 \cm, 
	but is known to be less accurate in higher wavenumber ranges 
	\cite{11HuScLe,11HuScLe2,12SuBrHuSc}. For example, there are 
	shifts in line position of up to 1.8 \cm\ around 6500 \cm\ 
	\cite{12SuBrHuSc}. Here we assess the accuracy of BYTe in the 
	range 7400 - 8600 \cm. Experimental vs theoretical stick spectra 
	at 21.5 $^{\circ}$C for the whole region is shown in 
	Figure~\ref{fig:byte1}. On the whole there is reasonable agreement, 
	although at high resolution BYTe line positions are shifted 
	from experimental ones by up to 3 \cm\ with an average of 1.6 \cm. 
	Overall BYTe intensities are within 20 - 55\%\ of the retrieved 
	experimental ones, through the difference is worse in a few cases. 
	Figure~\ref{fig:byte2} illustrates the comparison for the region 
	7800 -- 7900 \cm. Although the accuracy of BYTe is not enough to 
	make reliable assignments by line list comparison, there 
	are clear coincidences in strong lines (examples shown by symbols 
	in Figure~\ref{fig:byte2}) which gave us confidence that the 
	experimental \NH\ lines could be assigned using BYTe as a starting 
	point to initialise more rigorous methods.

	\subsection{Assignments}
	
	All assignments, vibrational bands and experimental energies 
	discussed below are new or observed for the first time in 
	this work.
	
	8468 experimental lines were retrieved in the region 
	7400 - 8600 \cm, in total 2474 of these have been 
	assigned, 1343 by combination differences (CD) and 
	1131 by the method of branches (Br). 2170 assigned lines have 
	unique upper and lower energy state quantum labels in BYTe. These are 
	fully assigned with 26 quantum labels, 13 per rotation-vibration state 
	as recommended by Down {\it et al} \cite{jt546}. The additional 304 do not 
	have unique upper state labels in BYTe hence only the rigorous quantum state 
	labels (total angular momentum and total symmetry) can be trusted. These are 
	partially assigned with 13 lower state labels, as recommended by Down {\it et al} 
	\cite{jt546}, and 2 upper state labels, $J$ and $\Gamma_{tot}$. The 5994
	unassigned experimental lines either did not have a CD partner or did 
	not correspond to a shifted BYTe line for any of the vibrational bands 
	analysed in this work. Most of these are weak lines that probably belong 
	to other vibrational bands, perhaps hot bands, that have too few observed 
	lines in the current experiment to be properly characterized. Around 1200 
	unassigned lines could be classified as strong or intermediate. These lines 
	may belong to other bands not included in the present analysis, else the assignment 
	of these lines may have been hampered by the uncertainty in experimental and 
	theoretical intensities. The experimental line list with full or partial quantum 
	labels for assigned lines is presented as supplementary data to this paper.
	
	Upper state energies were computed using MARVEL ground state energies 
	and the line position of the strongest line assigned to that state. 
	These are presented as supplementary data to this paper and summarised 
	in Table~\ref{tab:expE1} and Table~\ref{tab:expE2}. 
	
	Lines with full assignments belong to one of 15 different bands. 
	Table~\ref{tab:bands} gives a summary of the observed bands including the 
	number of lines assigned to each. The bands are listed in order of vibrational 
	band origin (VBO) which is either the computed upper energy for the observed P(0) 
	transition (E) or the BYTe prediction shifted by the average observed minus calculated 
	difference for the band (B). For simplicity abbreviated vibrational 
	labels $(v_1 v_2 v_3^{L_3} v_4^{L_4})^{i}$ are used to identify bands 
	in this table. $L_3$ and $L_4$ are vibrational angular momentum quantum 
	numbers corresponding to doubly degenerate $v_3$ and $v_4$ vibrational modes 
	respectively. $i$ is the inversion symmetry of the $v_2$ vibrational mode which 
	can be positive (+, symmetric) or negative (-, asymmetric). The full 26 quantum 
	labels for each transition, 13 per vibration-rotation state as recommended by 
	Down {\it et al} \cite{jt546}, are given in the supplementary data.
	
	Lines with partial assignments belong to one of 6 different bands. In the absence 
	of unique vibrational labels these bands are simply named Band 1 - 6. Line assignments 
	and derived energies for these bands, totalling 304 and 223 respectively, are summarised 
	in Table~\ref{tab:expE2}. Full lower and partial upper quantum labels for the assignments 
	are given in the supplementary data.  These bands are not the
first observed ammonia bands for which the vibrational quantum numbers 
have yet to be determined \cite{86CoLe,88LeCo}.
	
	The CD assignments can be divided into two categories, those where the calculated 
	upper energies agree within 0.006 \cm\ and those where the calculated upper energies 
	agree within 0.02 \cm. 
	
	The former can be considered secure, as the calculated upper energies consistently agree 
	within the higher experimental uncertainty for given CD pair and triplets or groups of 
	transitions sharing the same upper energy state.   
	
	In the latter case, the real transition(s) involved in the CD relation 
	are obscured in the experimental spectrum by (a) stronger line(s). In other words 
	the assigned line position is the nearest, rather than the true, peak-centre. 
	Hence these are more tentative.    
	
	The accuracy of Br assignments depends on the 
	determination of the Obs. - Calc. for a given vibrational band. 
	
	Strong bands with 10s of CD and Br assignments, such as 
	$(v_{2} + 2v_{3}^{2})^{-}$, are well characterised and the stability 
	of the Obs. - Calc. through the band can be clearly seen from the calculated 
	energies tabulated in the supplementary data. Hence lines assigned to these 
	bands by Br are reliable.  
	
	Weaker bands, such as $(2v_{1} + v_{4}^{1})^{-}$, are more tentative. Although 
	the overall consistency of our CD and Br results gives us confidence in all our 
	assignments. 
	
	\begin{table}
		\caption{Summary of fully assigned observed bands in order of vibrational band origin 
			(VBO) with abbreviated ($v_1 v_2 v_3^{L_3} v_4^{L_4} i$) vibrational labels. 
			$N_{lines}$ is the total number of lines assigned to the band. CD and Br are the 
			number of lines assigned using combination differences and the method of branches 
			respectively. The VBO has been derived from the observed P(0) transition (E) or BYTe (B).} 
		\begin{center}
			\begin{tabular}{llrrr}
				\hline
				Band                                    & VBO  / \cm     & $N_{lines}$ & CD & Br   \\
				\hline
				$(v_{1} + v_{2} + 2v_{4}^{2})^{+}$      & 7572.9549 E & 143 & 76 & 67    \\
				$(v_{1} + v_{2} + 2v_{4}^{2})^{-}$      & 7603.1713 E & 134 & 69 & 65    \\
				$(v_{1} + v_{2} + v_{3}^{1})^{+}$       & 7656.8700 E & 162 & 103 & 59    \\
				$(v_{1} + v_{2} + v_{3}^{1})^{-}$       & 7673.84~~ B & 135 & 72 & 63 \\
				$(v_{2} + 2v_{3}^{2})^{+}$              & 7854.3892 E & 145 & 101 & 44   \\
				$(v_{2} + 2v_{3}^{2})^{-}$              & 7864.0831 E & 152 & 106 & 46 \\
				$(2v_{1} + v_{4}^{1})^{+}$              & 8086.5926 E & 64 & 12 & 52    \\
				$(2v_{1} + v_{4}^{1})^{-}$              & 8089.59~~ B & 63 & 14 & 49    \\
				$(v_{1} + v_{3}^{1} + v_{4}^{1})^{+}$   & 8174.7017 E & 224 & 135 & 89   \\
				$(v_{1} + v_{3}^{1} + v_{4}^{1})^{-}$   & 8177.4358 E & 217 & 120 & 97   \\
				$(v_{1} + 3v_{4}^{1})^{+}$              & 8253.7494 E & 159 & 109 & 50   \\
				$(v_{1} + 3v_{4}^{1})^{-}$              & 8257.5341 E & 152 & 91 & 61   \\                      
				$(v_{1} + 2v_{2} + 2v_{4}^{2})^{+}$     & 8266.3284 E & 120 & 73 & 47  \\
				$(2v_{3}^{2} + v_{4}^{1})^{+}$          & 8463.2901 E & 151 & 54 & 97 \\
				$(2v_{3}^{2} + v_{4}^{1})^{-}$          & 8463.8719 E & 149 & 55 & 94 \\
				\hline
			\end{tabular} \label{tab:bands}
		\end{center}
	\end{table}
	
	\begin{table}
		\caption{Summary of fully assigned new \NH\ experimental energies above 7000 \cm\
			with abbreviated ($v_1 v_2 v_3^{L_3} v_4^{L_4} i$) vibrational labels 
			and maximum rotational quantum numbers $J_{\rm max}$ and $K_{\rm max}$. 
			$N_{E}$ is the total 
			number of experimentally derived energies for the band. 
			Obs. - Calc. gives the average experimental minus BYTe energy difference for the band in \cm.} 
		
		\begin{center}
			\begin{tabular}{lrrrr}
				\hline
				Band                                   & $N_{E}$ & $J_{\rm max}$ & $K_{\rm max}$ & Obs.-Calc.  \\
				\hline
				$(v_{1} + v_{2} + 2v_{4}^{2})^{+}$     & 104   &    14    &   14     & 2.3         \\
				$(v_{1} + v_{2} + 2v_{4}^{2})^{-}$     & 99   &    14    &   14     & 3.0         \\
				$(v_{1} + v_{2} + v_{3}^{1})^{+}$      & 101    &    16    &   16     & 0.6         \\
				$(v_{1} + v_{2} + v_{3}^{1})^{-}$      & 92    &    15    &   15     & 1.0         \\
				$(v_{2} + 2v_{3}^{2})^{+}$             & 85    &    15    &   15     & 2.9         \\ 
				$(v_{2} + 2v_{3}^{2})^{-}$             & 88    &    14    &   14     & 2.4         \\
				$(2v_{1} + v_{4}^{1})^{+}$             & 58    &    13    &    13     & 1.2         \\
				$(2v_{1} + v_{4}^{1})^{-}$             & 56    &    14    &    13    & 1.3        \\
				$(v_{1} + v_{3}^{1} + v_{4}^{1})^{+}$  & 151   &    15    &   15     & -1.9         \\
				$(v_{1} + v_{3}^{1} + v_{4}^{1})^{-}$  & 105   &    16    &   16     & -0.6         \\
				$(v_{1} + 3v_{4}^{1})^{+}$             & 101   &    14    &   14     & 2.3         \\
				$(v_{1} + 3v_{4}^{1})^{-}$             & 104   &    14    &   13     & 2.2         \\                     
				$(v_{1} + 2v_{2} + 2v_{4}^{2})^{+}$    & 81   &    11    &  11      & 1.5         \\
				$(2v_{3}^{2} + v_{4}^{1})^{+}$         & 123    &    13    &  13      & 2.4         \\
				$(2v_{3}^{2} + v_{4}^{1})^{-}$         & 121    &    12    &  12      & 2.2         \\
				\hline
			\end{tabular} \label{tab:expE1}
		\end{center}
	\end{table}
	
		\begin{table}
			\caption{Summary of partially assigned lines, and new \NH\ experimental energies above 7000 \cm\ 
				derived from them, with arbitrary band name and maximum rotational quantum numbers $J_{\rm max}$. 
				$N_{E}$ is the total number of experimentally derived energies for each band. $N_{lines}$ is the total 
				number of lines assigned to the band. CD and Br are the 
				number of lines assigned using combination differences and the method of branches 
				respectively. Obs. - Calc. gives the average experimental minus BYTe energy difference for the band in \cm. 
				Range gives the approximate span of the band in \cm.  The VBO has been derived from the observed P(0) transition (E) or BYTe (B).} 
			
			\begin{center}
				\begin{tabular}{clllrrrrr}
\hline					
Band	&	Range (approx.)	&VBO  / \cm     &$N_{lines}$	&	CD	&	Br	&	$N_{E}$	&	$J_{max}$&	Obs.-Calc. \\
\hline
1	&	7450 - 8000	& 7491.18~~B &	36	&	18	&	18	&	27	&	8	&	2.3 \\
2	&	8000 - 8450	& 8282.30~~B &	63	&	30	&	33	&	46	&	11	&	-1.9 \\
3	&	8000 - 8450	& 8283.5189 E &	69	&	24	&	45	&	56	&	11	&	-0.6 \\
4	&	8000 - 8400	& 8211.24~~B &	52	&	35	&	17	&	34	&	9	&	2.3 \\
5	&	8100 - 8400	& 8212.04~~B &	50	&	21	&	29	&	39	&	10	&	2.2 \\
6	&	8050 - 8400	& 8281.65~~B &	34	&	25	&	9	&	21	&	9	&	1.5 \\
\hline
				\end{tabular} \label{tab:expE2}
			\end{center}
		\end{table}
	
	\section{Summary}
	
	In this paper we present an experimental line list for, and an analysis of, 
	a 35 year old room temperature spectrum of \NH\ in the region 7400 -- 8600 \cm.
	The centers and intensities of 8468 ammonia lines were retrieved 
	using a multiline fitting of the spectrum. For isolated lines of intermediate 
	strength the accuracy of retrieved line position and intensities is estimated 
	to be of the order 3$\times10^{-3}$ \cm\ and 15 \%\ respectively. Although it should 
	be noted that the uncertainty in the retrieved intensity may significantly exceed 
	15\%\ for lines stronger that $1\times 10^{-22}$ cm/molecule. 
	
	A comparison between the measurements and BYTe shows in general good
	agreement but there are shifts in line position of up to 3 \cm\
	throughout the region and experimental line intensities
	are only reproduced with 20 - 55 \% at best.  Work towards a new, more accurate,
	hot \NH\ line list is currently being carried out as part of the
	ExoMol Project \cite{jt528}.
	
	The use of BYTe and MARVEL has allowed the assignment of 2474 
	lines, 1343 by combination differences and a further 
	1131 by the method of branches. In total 1692 
	new experimental energies between 7000 - 9000 \cm\ have been derived. 
	Assignments associated with strong bands with 
	tens of combination difference and branch assignments should all be 
	reliable, as these bands are well characterised and have stable 
	observed minus calculated differences throughout the band. The remaining 
	assignments should also be safe, as all results from the two assignment 
	procedures have proven very consistent, though these are more tentative.    
	
	We note that a room temperature spectrum for \NH\ in the
	9000 -- 10,000 \cm\ region, 
	also measured by Dr Catherine de Bergh in 1980, is available from the Kitt 
	Peak Archive. We plan to make this a focus of future work although it is 
	to be anticipated that current line lists will be less reliable at
	these higher wavenumbers. 
	
	\section*{Acknowledgements}
	
	This work was supported by a
	grant from Energinet.dk project N. 2013-1-1027, by
	UCL through the Impact Studentship Program and the European Research 
	Council under Advanced Investigator Project 267219.
	
\bibliographystyle{model1a-num-names}


\end{document}